\documentclass[english]{revtex4}
\usepackage[T1]{fontenc}
\usepackage[latin9]{inputenc}
\usepackage{amsmath}
\usepackage{amssymb}
\usepackage{graphicx}
\usepackage{esint}

\makeatletter
\@ifundefined{textcolor}{}
{%
 \definecolor{BLACK}{gray}{0}
 \definecolor{WHITE}{gray}{1}
 \definecolor{RED}{rgb}{1,0,0}
 \definecolor{GREEN}{rgb}{0,1,0}
 \definecolor{BLUE}{rgb}{0,0,1}
 \definecolor{CYAN}{cmyk}{1,0,0,0}
 \definecolor{MAGENTA}{cmyk}{0,1,0,0}
 \definecolor{YELLOW}{cmyk}{0,0,1,0}
}

\makeatother

\usepackage{babel}
\begin{document}

\title{Determinism and Quantum Mechanics with time Bell inequalities}

\author{Ramon Lapiedra}

\affiliation{Departament d'Astronomia i Astrofísica, Universitat de València,
46100 Burjassot, València, Spain\\
 }

\author{A. Pérez}

\affiliation{Departament de Física Teòrica and IFIC, Universitat de València-CSIC,
46100-Burjassot, València, Spain.\\
}
\begin{abstract}
We propose a definition of determinism for a physical system that
includes, besides the measurement device, the surrounding environment.
This \textit{enlarged system} is assumed to follow a predefined trajectory
starting from some (unknown) initial conditions that play the role
of hidden variables for the experiment. These assumptions, which are
different from realism, allow us to derive a type of time Bell inequalities,
which are violated by Quantum Mechanics. In order to illustrate this
violation, we discuss the particular case of measurements on a spin
1/2 particle.
\end{abstract}


\maketitle

\section{Introduction}

Let us consider the time evolution of an isolated physical system,
either classical or quantum. Does it always exist, as a matter of
principle, a trajectory which, from some initial conditions, gives
the values of the different quantities of the system during some finite
duration of time, as it is, for example, the case in Newtonian mechanics,
modulo some regularity conditions? This constitutes the concept of
determinism in nature (sensibly different from the concept of realism,
as we discuss later). As a matter of fact, determinism can be present
under some circumstances, but the problem we want to address is whether
determinism is \emph{always} present \emph{regardless of our capacity
of prediction in practice}. Therefore, to discard determinism it suffices
to find at least one example where it is contradicted by experimental
data. This is what we are going to discuss along the present paper.
Our claim is that determinism fails because it enters in contradiction
with Quantum Mechanics (QM) and, maybe, with experiments. 

Obviously QM, as described in standard textbooks, is an extremely
successful non-deterministic theory, since the result of measurements
can only be predicted in an statistical way. This is so even if the
equation of motion (the Schrödinger equation, for non relativistic
QM) provides the future state of the system when the initial conditions
are specified. In this sense, the time evolution of QM might be considered
as deterministic. It is the measurement process that introduces the
non deterministic nature of QM, since the result of a measurement
can not, in general, be predicted with certainty from the knowledge
of the state corresponding to the quantum state. Moreover, this state
is modified by the measurement, leading to the well known collapse
of the wave function. In other words, the measurement process is invasive
(in the precise sense described in \cite{PhysRevLett.54.857}) with
respect to the system to be measured. 

Thus, even if we admit that the quantum time evolution can be considered
as deterministic, the action of the measurement device on the system
under consideration appears as an extra ingredient that modifies the
conditions of that system. Once we accept this, the question arises:
Is it possible to postulate a deterministic evolution that describes,
not only the quantum system (represented by $Q$) under study, but
also the measurement apparatus and, if necessary, the surrounding
environment? We will refer to this system as the \textit{enlarged
system} (designated by $E$). Our deterministic hypothesis will be
formulated for this enlarged system, $E$, and not for the quantum
system, $Q$, to be measured. This has to be clearly formulated in
order to avoid further misunderstandings: The state of $Q$ may be
(and in general it will be) altered upon interaction with the measurement
device in an impredictable way. However, system E as a whole will
follow a predetermined trajectory that evolves from some initial conditions,
according to the deterministic hypothesis we want to be tested by
confrontation with the experiment.

The definition of determinism we introduced above differs from realism,
can be discarded by proposed experiments, and is based on testing
some \textit{time Bell-like} inequalities, which take the form

\begin{equation}
|P(a,b)-P(a,c)|\le1-P(b,c),\label{inequalityintro}
\end{equation}
 where $P(a,b)$ is the expected value associated to consecutive measurements
of spin directions $a$ and $b$ (similarly for the rest of magnitudes
on this equation), on the \textbf{same particle}.

As shown in the next Section, we can find some scenarios in which
determinism enters in contradiction with QM. The reason why the violation
of usual Bell inequalities can be reconciled with \textbf{non local}
realism, while our time Bell-like inequalities cannot be reconciled
with determinism, is that the deterministic assumption is a stronger
condition than realism. More precisely, realism postulates the existence
of some hidden variables values behind the outcome of a performed
measurement, \emph{without requiring that these hidden values remain
the same after performing the measurement}. On the contrary, the deterministic
assumption for the \textit{enlarged system} $E$ postulates the existence
of the same \emph{standing} hidden variable values (the unknown initial
conditions) behind \emph{all} the successive self-responses of $E$
along a certain finite time. As already mentioned, this \textit{enlarged
system} includes, at least the measurement device besides the measured
system $Q$.

An important ingredient in the experimental setup we introduce to
discuss the violation of the time Bell inequalities is the device
that makes the selection of the measurement directions. We would like
to decouple the choice of these directions from the rest of the measurement
apparatus as much as possible. For that reason, we assume that this
choice is performed by a (deterministic) pseudorandom generator that
has been previously designed (prior to the starting of the experiment)
and works independently of the measurement apparatus. As we will see
later, the introduction of this independent device plays a major role
in the derivation of our time Bell inequalities.

Similar results to the ones reported above have been previously obtained
by De Zela \cite{PhysRevA.76.042119} by \textit{adding} to the deterministic
postulate a \emph{non contextuality} condition, which amounts to saying
that initial conditions and measurement directions are uncorrelated.
But this condition becomes specially unjustified when determinism
is assumed, since then these directions depend on the initial conditions.
In the present paper, we propose a kind of experiment for which determinism
by itself does not entail contextuality.

\section{Determinism and the violation of time Bell-like inequalities}

We start with a large ensemble of free spin $1/2$ particles numbered
by $n=1,2,3,\cdots N$, all of them prepared on the same quantum state,
and consider the following ideal experiment: we first fix a set of
three space directions given by the unit 3-vectors $\vec{a}$, $\vec{b}$
and $\vec{c}$. Then, on each particle we perform two successive spin
measurements along two randomly selected directions out of this set:
We prepare one particle and make two consecutive measurements, then
prepare the next particle, reset time to zero and proceed in the same
way, etc ...

These \textquotedbl{}free\textquotedbl{} particles could evolve interacting,
to some extent, with their environment in an uncontrollable way, and
will certainly interact with the measurement apparatus. The latter
includes the device that randomly selects the two measurement directions
out of the three initially fixed directions. Consider now the physical
system that includes the particles to be measured, the experimental
facility and the interacting environment. In accordance with the Introduction,
we will refer to this system as the \textit{enlarged system}, hereafter
represented by $E$, and will assume that such system can be considered
as an isolated system during the whole experiment. 

We now define our notion of determinism following what happens, for
example, with Newtonian determinism (where, aside from some ``pathological''
cases, the initial position and velocity, i. e., the initial conditions,
allow us to know some piece of the trajectory of a particle). We will
make the corresponding deterministic hypothesis \textbf{for the enlarged
system} $E$. Then, not only the successive spin measurement outcomes,
$\pm\hbar$/2, are determined from the initial conditions on $E$,
but also the successively selected measurement directions are determined
too, whatever the selection mechanism is: The precise way this will
be implemented is discussed below. Notice that we do not impose any
restrictions on the assumed initial conditions: In particular, they
could be non local, i.e. they could range over non causally connected
space-time regions.
\begin{figure}
\includegraphics[width=8cm]{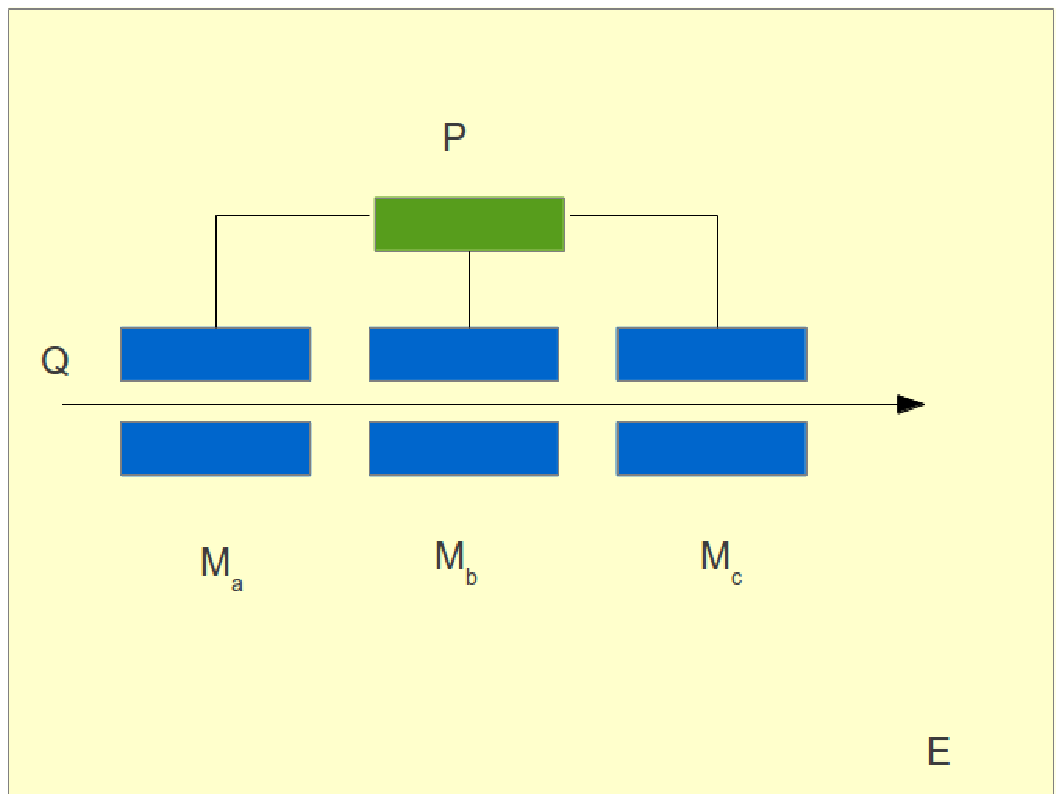}

\caption{A sketch of the measurement apparatus, as described in the text. $M_{a}$,
$M_{b}$ and $M_{c}$ perform measurements on the system $Q$ (a spin
1/2 particle) according to the selection made by the deterministic
device $P$. Only two consecutive measurements are performed. The
experimental setup is contained on some enlarged system $E$, which
is assumed to be isolated.}

\end{figure}

Let us be more precise about our enlarged system and the measurement
process, which is sketched in Fig. 1. We denote by $\lambda$ the
initial conditions that we postulate to exist, complementary to the
prepared state of the quantum description, but leading to the same
statistical predictions than this quantum description. These initial
conditions will belong not to the particle, but to the whole system
$E$. Imagine that, each time we prepare our spin 1/2 particles, system
$E$ starts from different initial conditions, i. e., different $\lambda$
values. We will perform two consecutive spin measurements on each
prepared particle. These two consecutive spin measurements will be
performed at two randomly selected times, out of three fixed values
$t_{1},t_{2}$, and $t_{3}$ relative to the preparation time (which
is always reset to zero). To each selected time, $t_{1}$, $t_{2}$
or $t_{3}$, we associate a constant measurement direction, $\vec{a}$,
$\vec{b}$ and $\vec{c}$, respectively. In other words, we originally
establish a given correspondence $\{t_{1}\rightarrow\vec{a},t_{2}\rightarrow\vec{b},t_{3}\rightarrow\vec{c}\}$
and keep it unchanged during the measurement process for the entire
set of particles. Thus, the randomness in the direction selection
arises only as a consequence of the randomness in the selected pair
of times out of the set $\{t_{1},t_{2},t_{3}\}$. The device (hereafter
referred to as $P$) that selects the pair of times, $(t_{1},t_{2})$,
$(t_{1},t_{3})$ or $(t_{2},t_{3})$ (and so the choice of the pairs
of measurement directions) could, in fact, be a pseudorandom generator,
as long as the statistics it provides for this selection is close
to a true random one. In the present paper, it is assumed to be previously
manufactured (it could simply consist on a deterministic pseudorandom
computer program), and is included into the experimental setup prior
to the starting of the experiment: prior to the entire succession
of measurement pairs. It will start producing the mentioned pairs
from a ``seed'' defined by a set of parameters which we represent
by the initial condition $\Lambda$. Notice, then, that our device
$P$ can be designed, in particular, in such a way that the series
of pseudorandom choices is fixed, i.e., if we repeat the measurement
with a new set of $N$ particles, while the starting $\Lambda$ is
the same as for the previous set, $P$ will produce exactly the same
series of direction choices. The introduction of this deterministic
selection device, which is working independently of the rest of system
$E$, will be a crucial point all along the present paper.

Then, let us denote by $S$ the values of the particle measurement
outcomes, which are conveniently normalized to $\pm1$. According
to the determinism postulate for our enlarged system, there exists
a function (unknown to us) which provides those outcomes for each
value of time $t_{i}$, starting from the initial conditions, i. e.,
the parameter values $\lambda$. Let us represent this function by
$S=S(\lambda,t_{i},\vec{x}(t_{i}))$, $i=1,2,3$ , with $\vec{x}(t_{i})\in\{\vec{a},\vec{b},\vec{c}\}$.
Notice that this notation for the function $S$ is actually redundant:
according to the above discussion, once we have fixed $\lambda$ and
$t_{i}$, the value of $S$ becomes determined, therefore we could
drop the argument $\vec{x}$ in $S$, although we will keep it for
convenience in the following discussion.

We now follow the original proof of usual Bell inequalities \cite{Bell:1964kc}
in order to arrive to similar inequalities for our consecutive measurement
outcomes. Let us consider the following three expectation values:

\begin{equation}
P(a,b)=\int d\lambda\rho(\lambda)S(\lambda,t_{1},\vec{a})S(\lambda,t_{2},\vec{b}),\label{expectvaluab}
\end{equation}

\begin{equation}
P(a,c)=\int d\lambda\rho(\lambda)S(\lambda,t_{1},\vec{a})S(\lambda,t_{3},\vec{c}),\label{expectvaluac}
\end{equation}

\begin{equation}
P(b,c)=\int d\lambda\rho(\lambda)S(\lambda,t_{2},\vec{b})S(\lambda,t_{3},\vec{c}),\label{expectvalubc}
\end{equation}
 where $\rho(\lambda)$ stands for a the probability distribution
of the $\lambda$ values, which satisfies $\int d\lambda\rho(\lambda)=1$.
There are some subtelties related to the derivation of the above equations.
The interested reader is addressed to the Appendix. Let us go on with
the derivation of our time Bell-like inequalities. We take the difference

\begin{align}
P(a,b)-P(a,c) & =\int d\lambda\rho(\lambda)S(\lambda,t_{1},\vec{a})\label{difference}\\
\times & [S(\lambda,t_{2},\vec{b})-S(\lambda,t_{3},\vec{c})].
\end{align}

Henceforth, the proof of the inequalities goes along the same lines
as the proof of the original Bell inequalities in Bell's seminal paper
\cite{Bell:1964kc}. First, since $S^{2}(\lambda,t_{2},\vec{b})=1$,
the above difference can be written as

\begin{eqnarray}
P(a,b)-P(a,c) & = & \int d\lambda\,\rho(\lambda)S(\lambda,t_{1},\vec{a})S(\lambda,t_{2},\vec{b})\nonumber \\
 &  & \,\,\,[1-S(\lambda,t_{2},\vec{b})S(\lambda,t_{3},\vec{c})].\label{difference2}
\end{eqnarray}

Then, taking absolute values, we are led to

\begin{equation}
|P(a,b)-P(a,c)|\le\int d\lambda\rho(\lambda)[1-S(\lambda,t_{2},\vec{b})S(\lambda,t_{3},\vec{c})],\label{inequality}
\end{equation}
 that is, to the well known Bell inequality

\begin{equation}
|P(a,b)-P(a,c)|\le1-P(b,c),\label{inequality2}
\end{equation}
 now referring to two consecutive measurements on the \textbf{same
particle}.

In QM, leaving aside the experimental difficulties to perform the
kind of experiment we are considering (see \cite{PhysRevA.76.042119}
for some sound proposals), the three mean values in (\ref{inequality2})
can be theoretically calculated as the corresponding expected values.
These values become $P(a,b)=\vec{a}.\vec{b}$, irrespective of the
spin particle state prior to the first measurement (see for instance
\cite{Lapiedra09}) and similarly for $P(b,c)$ and $P(a,c)$. Thus,
inequality (\ref{inequality2}) becomes 
\begin{equation}
|\vec{a}.\vec{b}-\vec{a}.\vec{c}|+\vec{b}.\vec{c}\le1,\label{inequality3}
\end{equation}
 which is violated for $\vec{b}.\vec{c}=0$ and $\vec{a}=(\vec{b}-\vec{c})/\sqrt{2}$,
in which case the left hand side of inequality (\ref{inequality3})
reaches the value $\sqrt{2}$.

Thus, for the enlarged system consisting on 1/2-spin particles and
the measurement apparatus, including the device selecting, once forever,
the time pairs and the corresponding measurement directions, plus
the affecting environment if any, the assumed determinism enters in
contradiction with quantum mechanics. In other words, the essence
of the present paper is the following: could we consider \textbf{any
system} $E$ larger than the particle under study, so as to encompass
the measurement device, perhaps the laboratory and even beyond, in
order to attain an enlarged system whose evolution fulfills the deterministic
assumption? Our claim is that, if the appropriate measurements were
performed and inequality (\ref{inequality2}) was found to be violated,
as we expect from QM, then, no matter how large $E$ was assumed to
be, the answer to this question would be negative.

At this point, it is interesting to compare our result with a similar
one stated in \cite{0034-4885-71-2-022001}: there, under the following
three postulates, \emph{macroscopic realism per se}, \emph{noninvasive
measurability} and \emph{induction}, Leggett derives some Clauser-Horn-Shimony-Holt
(CHSH) inequalities \cite{PhysRevLett.23.880,refId0}, relating the
successive outcomes of a system with random dichotomic responses against
four types of measurements. Our determinism postulate entails macroscopic
realism per se and the induction postulate, and also noninvasive measurability
\emph{for the enlarged system} (the spin 1/2 particles plus the environment
including the measurement device). Thus, a difference between Leggett's
approach and ours is that we replace the dubious non-invasive measurability
for the measured system $Q$ by a more clear postulate as determinism
for the isolated enlarged system, which does not entail noninvasive
measurability for the measured system. Since determinism is assumed
for the whole system $E$, this implies that all variables in this
system, including the system being monitored, are defined at all times
once some starting initial conditions $\lambda$ are specified (even
if unknown). Non-invasive measurability for the spin 1/2 particle
would mean that a measurement performed at time $t_{1}$ will not
affect a measurement at time $t_{2}$. Instead, determinism implies
that measurements made at times $t_{1}$ \textbf{and} $t_{2}$ are
predetermined by some initial conditions, at the cost that these conditions
belong to the whole system $E$. We recall here that noninvasive measurability
has been criticized thoroughly in the literature. More recently, this
hypothesis has been discussed and criticized in \cite{Zukowski2010}.

Let us summarize the hypothesis we introduced in order to derive our
time Bell-like inequalities (\ref{inequality2}), based on determinism,
as compared to other hypothesis currently made in similar areas: 
\begin{itemize}
\item We assume \textit{determinism} defined as follows: One can define
an enlarged isolated system $E$ that includes the particles to be
measured, the experimental setup and the surrounding environment,
such that all variables in that system evolve deterministically, from
some unknown initial conditions $\lambda$. This amounts to assuming
noninvasive measurability, but for the whole system $E$, and \textbf{not
for the particles we are measuring}. Determinism implies \textit{causality}:
Initial conditions are not affected by measurements performed later
in time. 
\item We do not need to assume\textit{ freedom (or statistical independence})
defined as follows: The joint distribution of values for the initial
conditions $\lambda$ of the enlarged system and those providing the
measuring directions factorizes. In our setup, this independence is
warrantied by the nature of the device $P$, which has been previously
manufactured and evolves deterministically from its own initial condition
$\Lambda.$ Notice that statistical independence corresponds, in our
case, to the above mentioned \textit{non-contextuality}, that we did
not need to assume. 
\item We do not need to assume \textit{reproducibility,} as an extra hypothesis,
defined as follows: The measured expecting values reach standing values
as the number of measurements increases beyond some definite given
level. Reproducibility may or not apply to a given system. In the
considered case of a spin 1/2 particle \textit{reproducibility} is
experimentally present and we would reach the standing values predicted
by QM.
\end{itemize}

\section*{Conclusions}

The results obtained in the present paper can be summarized as follows:
Modulo the absence of any \textit{conspiracy} \cite{refId0} and without
any other restriction, we have proved that determinism for whatever
\emph{enlarged} system containing a successively measured 1/2 spin,
the measurement device and any possibly interacting environment, is
in contradiction with QM. It is to be remarked that the same absence
of conspiracy is seen to be nedeed when proving the original Bell
inequalities \cite{Bell:1964kc} and other similar inequalities. Thus,
either Quantum Mechanics, or determinism (as defined here), must be
false. So, if we accept QM, in view of its great success we must conclude
that determinism would contradict experiments. 
\begin{acknowledgments}
This work has been supported by the Spanish Ministerio de Educación
e Innovación, MICIN-FEDER project No. FIS2012-33582, and by Projects
FPA2011-23897 and ``Generalitat Valenciana'' grant PROMETEO/2009/128.
R. L. also thanks Dr Michael Hall for having made a number of useful
critical comments to the manuscript, and Prof. Eliseo Borrás for his
reading and comments. Useful discussions with A. Palacios-Laloy are
gratefully recognized. 
\end{acknowledgments}

\section*{Appendix}

As we show below, in order to arrive to Eqs. (\ref{expectvaluab}-\ref{expectvalubc}),
we need to have in mind that the selection of the successive direction
pairs is performed according to some deterministic device, $P$ (working
as a pseudo-random generator). The reason why we will need it in order
to arrive to Eqs. (\ref{expectvaluab}-\ref{expectvalubc}) is that
our deterministic system dynamics could correlate, say, the pair $\{\vec{a},\vec{b}\}$
of measurement directions to a set of particular values of the $\lambda$
variables. We name this hypothetical subset $\{\lambda_{ab}\}$, and
similarly for $\{\lambda_{ac}\}$ and $\{\lambda_{bc}\}$, respectively.
Consequently, in an evident notation, we would have three different
probability distributions, $\rho_{ab}$, $\rho_{ac}$ and $\rho_{bc}$,
instead a unique one, $\rho(\lambda)$, as in Eqs. (\ref{expectvaluab}-\ref{expectvalubc}).
Following these ideas, one then should write

\begin{equation}
P(a,b)=\int d\lambda\rho_{ab}(\lambda)S(\lambda,t_{1},\vec{a})S(\lambda,t_{2},\vec{b}).\label{eq2}
\end{equation}

\begin{equation}
P(a,c)=\int d\lambda\rho_{ac}(\lambda)S(\lambda,t_{1},\vec{a})S(\lambda,t_{3},\vec{c}).\label{eq2-1}
\end{equation}

\begin{equation}
P(b,c)=\int d\lambda\rho_{bc}(\lambda)S(\lambda,t_{2},\vec{b})S(\lambda,t_{3},\vec{c}),\label{eq2-2}
\end{equation}
instead of (\ref{expectvaluab})-(\ref{expectvalubc}), and we could
not conclude the proof of the time Bell inequalities (\ref{inequality2}).
In all, we would have three different probability spaces (one for
each subset $\{\lambda_{ab}\}$, $\{\lambda_{ac}\}$, $\{\lambda_{bc}\}$)
instead of a common probability space for the three expectation values
$P(a,b)$, $P(a,c)$, $P(b,c)$.

But, can the set of initial conditions, $\{\lambda\}$, really split
into three, perhaps overlapping, subsets $\{\lambda_{ab}\}$, $\{\lambda_{ac}\}$
and $\{\lambda_{bc}\}$ (with corresponding probability distributions),
as a consequence of the deterministic assumption, or can we alternatively
assume \emph{non contextuality}, that is $\{\lambda_{ab}\}=\{\lambda_{ac}\}=\{\lambda_{bc}\}\equiv\{\lambda\}$
with a common probability distribution, despite determinism? Let us
see why, in our case, determinism by itself does not imply \emph{contextuality},
i.e. it does not imply $\{\lambda_{ab}\}\ne\{\lambda_{ac}\}\ne\{\lambda_{bc}\}$:

1) First, remember that the two consecutive measurement directions
of the pair $(\vec{x},\vec{y})$ on the $n$th particle are applied
at the consecutive corresponding times, say $t_{x}$ and $t_{y}$,
counted from the corresponding initial time $t_{n}$, reset to zero
for each particle $n$. To this initial time, $t_{n}$, it corresponds
some initial condition $\lambda$, say $\lambda_{n}$. Now, let us
invoke the basic ``principle of causality'', according to which
the future cannot affect the past, which is actually included in our
deterministic hypothesis. Then, an action (the $\vec{x}$ or the $\vec{y}$
measurements on the particle) that is performed at $t_{x}$ or $t_{y}$,
respectively, that is, after the initial time, cannot influence what
was present for the particle in this previous initial time: In other
words, it cannot influence the corresponding initial condition $\lambda_{n}$.
Thus, as we wanted to show, this kind of action could never lead to
the above hypothetical splitting.

2) However, this is by no means the end of the story, since perhaps
things could go the other way around: we mean that, perhaps each $\lambda_{n}$,
let us say $\lambda_{n(xy)}\equiv\lambda_{xy}$ is responsible, at
least partially, for the ulterior two consecutive measurements being
just performed in the $(\vec{x},\vec{y})$ directions, thus finally
splitting the set $\{\lambda\}$ into the three subsets $\{\lambda_{ab}\}$,
$\{\lambda_{ac}\}$ and $\{\lambda_{bc}\}$. But this loophole cannot
hold since, by construction (by the deterministic character of $P$),
$\vec{x}$ and $\vec{y}$ are functions of time and of the UNIQUE
initial condition $\Lambda$, and so cannot depend on $\lambda$.
In other words, the selecting device $P$ and the initial conditions
$\lambda$ are, by construction, \textbf{uncorrelated} in our setup.
One might replace the role played by this device in choosing the measuring
directions by the hypothesis of ``freedom'' in making that choice,
so as to warrant that the joint distribution of values for $\lambda$
and $\Lambda$ factorizes \cite{Zukowski2010}. Here we do not need
to introduce such a hypothesis since this factorization appears naturally,
as a consequence of the assumption that our deterministic device $P$
has been previously designed and then included into the experiment,
therefore it works independently of the other elements of the full
system.

All in all: the different kinds of measurement actions cannot be correlated
with the different initial conditions $\lambda$, as we wanted to
prove.

There is still, in the above reasoning, a subtle difficulty to comment:
perhaps the different initial conditions $\lambda$ never repeat completely
themselves, such that the three hypothetical subsets, $\{\lambda_{ab}\}$,
$\{\lambda_{ac}\}$ and $\{\lambda_{bc}\}$, cannot achieve an standing
equality $\{\lambda_{ab}\}=\{\lambda_{ac}\}=\{\lambda_{bc}\}$, which
would prevent us from using the above argument to prove that determinism,
by itself, does not lead to contextuality. Nevertheless, this hypothetical
inequality would not be a problem, since what is physically relevant
are not the, for example, $\lambda_{ab}$ values by themselves, but
the corresponding function values $S(\lambda_{ab},t_{1})$, entering
in the expectation value $P(a,b)$, in (\ref{eq2}). Then, remember
the ''reproducibility'' of the measurement outcomes, meaning by
this that, in accordance to QM, those expectation values reach standing
values for a sufficiently large number, $N$, of measurement pairs.
This implies that, beyond $N$, we still could formally write $P(a,b)$
in (\ref{eq2}) using the same finite subset, $\{\lambda_{ab}\}$,
that we have used just for $N$. Thus, in practice, because this ``reproducibility''
principle, (which, in the present case, can be taken as an experimental
fact) we can use a finite number of $\lambda_{ab}$ values, and the
same for the $\lambda_{ac}$ and the $\lambda_{bc}$, to avoid our
initial difficulty. As a result, we still can use the above argument
in favor of non contextuality in spite of the deterministic assumption.

But, even admitting the impossibility of the above splitting action
driven by the deterministic postulate, it could be still possible
that the dynamics of system $E$ is originally arranged in such a
way that the values of the $\lambda$ variables are \textit{effectively}
correlated with the measurement directions because of the above possibility,
raised by Bell, of an initial \emph{conspiracy}: Hence we necessarily
would have to write (\ref{eq2}-\ref{eq2-2}), instead of (\ref{expectvaluab}-\ref{expectvalubc}),
in the absence of a common probability space. One can, in fact, \textit{design}
a \textit{deterministic device} (or a computer program), in the sense
that determinism is defined in the present paper, that incorporates
\textit{ad hoc} such correlations, even the kind of correlations needed
to simulate QM, under the form of convenient distribution functions
$\rho_{ab}$, $\rho_{ac}$ and $\rho_{bc}$. Consequently, it is easy
to see that one could not go on and obtain the desired time Bell-like
inequalities in such a case. At this point, we have to stress that
we are in a similar position to \textbf{any} derivation of usual Bell-like
inequalities, including the ones derived in \cite{PhysRevA.76.042119},
where this \emph{conspiracy} has always to be ruled out, in order
to complete any of these derivations.

Let us come back once more to the ''reproducibility'' question.
In the experimental setup described above, Eqs. (\ref{expectvaluab}-\ref{expectvalubc})
are understood as an statistical average over a large number of particles.
Each pair of measurements is in correspondence, within our scheme,
with some value $\lambda$. A comparison of these formulae with the
experimental results only makes sense if the set of $\lambda$ values
``accessed'' by the whole experiment is assumed to be \textit{statistically
representative} of the whole set. Of course, in our case, where the
\emph{enlarged} system, $E$, includes the experimental set up and
beyond, one might argue that the space of hidden variables is so large
that one could never achieve a representative subset of the total,
in which case no comparison with the experiment would be possible.
However, in spite of this large size, we have, according to QM, well
defined expectation values $P(x,y)$. Thus, we can assume that, after
a sufficiently large number of measurements, we achieve some sort
of saturation for expressions Eqs. (\ref{expectvaluab}-\ref{expectvalubc}).
That is, a further increase in the number of measurements will not
significantly change the result and the same experiment, repeated
under different initial conditions, will reproduce the same results:
We insist, this is just what happens with our setup according to QM
predictions. 

We think that it is important to have made explicit the postulate
of ruling out the \emph{conspiracy}, the observed (at least in accordance
with QM) ''reproducibility'' of the measurement outcomes, and the
incorporation of the deterministic $P$ device to select the successive
measurement directions, as prerequisites to arrive to Eqs. (\ref{expectvaluab}-\ref{expectvalubc})
in order to test determinism. It is important because, otherwise,
the reader could object from the very beginning that we should never
be able to find a contradiction between experiments and unrestricted
determinism, and then to conclude that something has had to go wrong
in the present paper with the claimed prediction of such a contradiction.
The reader might rise this objection because it seems intuitive that
unrestricted determinism could always be initially arranged so as
to fulfill any subsequent behavior of the system.

In all, while the derivation of usual Bell inequalities assumes local
realism, we have used here determinism to derive our time Bell-like
inequalities. Nevertheless, in both cases we need to rule out any
\emph{conspiracy}. A natural assumption that we have taken for granted,
in accordance to what is the usual practice in the literature on the
subject.

\bibliographystyle{apsrev}
\bibliography{/home/perez/investig/qcomputing/biblio/Bell}

\begin{thebibliography}{8}
\expandafter\ifx\csname natexlab\endcsname\relax\def\natexlab#1{#1}\fi
\expandafter\ifx\csname bibnamefont\endcsname\relax
  \def\bibnamefont#1{#1}\fi
\expandafter\ifx\csname bibfnamefont\endcsname\relax
  \def\bibfnamefont#1{#1}\fi
\expandafter\ifx\csname citenamefont\endcsname\relax
  \def\citenamefont#1{#1}\fi
\expandafter\ifx\csname url\endcsname\relax
  \def\url#1{\texttt{#1}}\fi
\expandafter\ifx\csname urlprefix\endcsname\relax\def\urlprefix{URL }\fi
\providecommand{\bibinfo}[2]{#2}
\providecommand{\eprint}[2][]{\url{#2}}

\bibitem[{\citenamefont{Leggett and Garg}(1985)}]{PhysRevLett.54.857}
\bibinfo{author}{\bibfnamefont{A.~J.} \bibnamefont{Leggett}} \bibnamefont{and}
  \bibinfo{author}{\bibfnamefont{A.}~\bibnamefont{Garg}},
  \bibinfo{journal}{Phys. Rev. Lett.} \textbf{\bibinfo{volume}{54}},
  \bibinfo{pages}{857} (\bibinfo{year}{1985}).

\bibitem[{\citenamefont{De~Zela}(2007)}]{PhysRevA.76.042119}
\bibinfo{author}{\bibfnamefont{F.}~\bibnamefont{De~Zela}},
  \bibinfo{journal}{Phys. Rev. A} \textbf{\bibinfo{volume}{76}},
  \bibinfo{pages}{042119} (\bibinfo{year}{2007}).

\bibitem[{\citenamefont{Bell}(1964)}]{Bell:1964kc}
\bibinfo{author}{\bibfnamefont{J.~S.} \bibnamefont{Bell}},
  \bibinfo{journal}{Physics} \textbf{\bibinfo{volume}{1}}, \bibinfo{pages}{195}
  (\bibinfo{year}{1964}).

\bibitem[{\citenamefont{Lapiedra}(2006)}]{Lapiedra09}
\bibinfo{author}{\bibfnamefont{R.}~\bibnamefont{Lapiedra}},
  \bibinfo{journal}{Europhys. Lett.} \textbf{\bibinfo{volume}{75}},
  \bibinfo{pages}{2002} (\bibinfo{year}{2006}), \eprint{quant-ph/0601185}.

\bibitem[{\citenamefont{Leggett}(2008)}]{0034-4885-71-2-022001}
\bibinfo{author}{\bibfnamefont{A.~J.} \bibnamefont{Leggett}},
  \bibinfo{journal}{Reports on Progress in Physics}
  \textbf{\bibinfo{volume}{71}}, \bibinfo{pages}{022001}
  (\bibinfo{year}{2008}).

\bibitem[{\citenamefont{Clauser et~al.}(1969)\citenamefont{Clauser, Horne,
  Shimony, and Holt}}]{PhysRevLett.23.880}
\bibinfo{author}{\bibfnamefont{J.~F.} \bibnamefont{Clauser}},
  \bibinfo{author}{\bibfnamefont{M.~A.} \bibnamefont{Horne}},
  \bibinfo{author}{\bibfnamefont{A.}~\bibnamefont{Shimony}}, \bibnamefont{and}
  \bibinfo{author}{\bibfnamefont{R.~A.} \bibnamefont{Holt}},
  \bibinfo{journal}{Phys. Rev. Lett.} \textbf{\bibinfo{volume}{23}},
  \bibinfo{pages}{880} (\bibinfo{year}{1969}).

\bibitem[{\citenamefont{{Bell, J. S.}}(1981)}]{refId0}
\bibinfo{author}{\bibnamefont{{Bell, J. S.}}}, \bibinfo{journal}{J. Phys.
  Colloques} \textbf{\bibinfo{volume}{42}}, \bibinfo{pages}{C2}
  (\bibinfo{year}{1981}).

\bibitem[{\citenamefont{Zukowski}(2010)}]{Zukowski2010}
\bibinfo{author}{\bibfnamefont{M.}~\bibnamefont{Zukowski}}
  (\bibinfo{year}{2010}), \eprint{1009.1749}.

\end{thebibliography}

\end{document}